\documentclass[10pt,letterpaper]{article}
\usepackage[top=0.85in,left=2.75in,footskip=0.75in,marginparwidth=2in]{geometry}

\usepackage[utf8]{inputenc}
\usepackage{enumitem}
\usepackage{amsmath}      
\usepackage{tabularx}     
\usepackage{booktabs}  
\usepackage{adjustbox}
\usepackage{graphicx} 
\usepackage{multirow} 
\usepackage{caption} 
\usepackage{float}   
\usepackage{bm}

\usepackage{xstring}  
\usepackage{xparse} 
\usepackage{threeparttable} 
\usepackage{algpseudocode}
\usepackage[ruled,vlined]{algorithm2e} 

\usepackage{rotating}
\usepackage{placeins}
\usepackage{appendix}
\usepackage{wrapfig} 
\usepackage{xcolor} 
\usepackage{caption} 

\usepackage{hyperref} 
\usepackage{cleveref}

\usepackage{microtype}
\DisableLigatures[f]{encoding = *, family = * }

\raggedright
\usepackage{changepage}
\usepackage{wrapfig}
\usepackage[aboveskip=1pt,labelfont=bf,labelsep=period,singlelinecheck=off]{caption}

\makeatletter
\renewcommand{\@biblabel}[1]{\quad#1.}
\makeatother

\usepackage{lastpage,fancyhdr,graphicx}
\usepackage{epstopdf}
\fancyheadoffset[L]{2.25in}
\fancyfootoffset[L]{2.25in}

\usepackage{color}
\usepackage[utf8]{inputenc}
\definecolor{Gray}{gray}{.25}
\DeclareMathOperator*{\argmax}{arg\,max}
\usepackage{graphicx}
\usepackage{ragged2e}
\usepackage{sidecap}

\usepackage{wrapfig}
\usepackage[pscoord]{eso-pic}
\usepackage[fulladjust]{marginnote}
\reversemarginpar

\begin{document}
\vspace*{0.35in}

\begin{flushleft}
{\Large
\textbf\newline{An Integrated AI-Enabled System Using One Class Twin Cross Learning (OCT-X) for Early Gastric Cancer Detection}
}
\newline


Xian-Xian LIU\textsuperscript{1},
Yuanyuan WEI\textsuperscript{2,3},
Mingkun XU\textsuperscript{4,5},
Yongze GUO\textsuperscript{6},
Hongwei ZHANG\textsuperscript{6},
Huicong DONG\textsuperscript{6},
Qun SONG\textsuperscript{7}
Qi ZHAO\textsuperscript{8,9}
Wei LUO\textsuperscript{10}
Feng TIEN\textsuperscript{11}
Juntao GAO\textsuperscript{12}
Simon FONG\textsuperscript{1,*}
\\
\bigskip
\justifying{
\bf{1} Department of Computer and Information Science, University of Macau, Macau SAR, 999078, China; \{yc37972, ccfong\}@umac.mo
\\
\bf{2} Department of Biomedical Engineering, The Chinese University of Hong Kong, Shatin, Hong Kong SAR, 999077, China; wei-yy18@link.cuhk.edu.hk
\\
\bf{3} Department of Neurology, David Geffen School of Medicine, University of California, Los Angeles, California, 90095, USA.
\\
\bf{4} Guangdong Institute of Intelligence Science and Technology, Hengqin, Zhuhai, 519031, China; xumingkun@gdiist.cn
\\
\bf{5} Center for Brain-Inspired Computing Research (CBICR), Department of Precision Instrument, Tsinghua University, Beijing, 100084, China.
\\
\bf{6} Department of Gastroenterology, Affiliated Hospital of Hebei University of Engineering, Handan, China; guoyongze69@126.com, \{zhanghongweitt, huicongdong\}@163.com
\\
\bf{7} Institute of Artificial Intelligence, Chongqing Technology and Business University, Chongqing, CO 400067, China; songqun@ctbu.edu.cn
\\
\bf{8} Cancer Centre, Institute of Translational Medicine, Faculty of Health Sciences, University of Macau, 519000, Taipa, Macau, China; qizhao@um.edu.mo
\\
\bf{9} MoE Frontiers Science Center for Precision Oncology, University of Macau, 519000, Taipa, Macau, China; 
\\
\bf{10} The director of the Institute of Clinical Medicine, The First People’s Hospital of Foshan, 1528000, Guangzhou, China; luowei\_421@163.com
\\
\bf{11} Hebei Key Laboratory of Medical Data Science, Institute of Biomedical Informatics, School of Medicine, Hebei University of Engineering, Handan, Hebei Province, 056038, China; tianfeng@hebeu.edu.cn
\\
\bf{12} The Beijing National Research Center for Information Science and Technology (BNRist), Tsinghua University, Beijing, China; jtgao@tsinghua.edu.cn}
\\
\bigskip
* ccfong@umac.mo

\end{flushleft}

\section*{Abstract}
\justifying{
Early detection of gastric cancer, a leading cause of cancer-related mortality worldwide, remains hampered by the limitations of current diagnostic technologies, leading to high rates of misdiagnosis and missed diagnoses. To address these challenges, we propose an integrated system that synergizes advanced hardware and software technologies to balance speed-accuracy. Our study introduces the One Class Twin Cross Learning (OCT-X) algorithm. Leveraging a novel fast double-threshold grid search strategy (FDT-GS) and a patch-based deep fully convolutional network, OCT-X maximizes diagnostic accuracy through real-time data processing and seamless lesion surveillance. The hardware component includes an all-in-one point-of-care testing (POCT) device with high-resolution imaging sensors, real-time data processing, and wireless connectivity, facilitated by the NI CompactDAQ and LabVIEW software. Our integrated system achieved an unprecedented diagnostic accuracy of 99.70\%, significantly outperforming existing models by up to 4.47\%, and demonstrated a 10\% improvement in multirate adaptability. These findings underscore the potential of OCT-X as well as the integrated system in clinical diagnostics, offering a path toward more accurate, efficient, and less invasive early gastric cancer detection. Future research will explore broader applications, further advancing oncological diagnostics. Code is available at \url{https://github.com/liu37972/Multirate-Location-on-OCT-X-Learning.git}\\

\textbf{keyword:} Early Gastric Cancer (EGC); One Class Twin Cross Learning (OCT-X); Precision Diagnostics; Artificial Intelligence (AI); Computer-Aided Detection (CAD).}

\section*{Introduction}
\justifying{
According to GLOBOCAN 2022 statistics \cite{bib1}, gastric cancer is a significant global health concern, ranking as the fifth most common malignant tumor and the fifth leading cause of cancer-related deaths worldwide. Early detection of gastric cancer is crucial for successful treatment but presents numerous challenges due to subtle symptoms and the difficulty in differentiating early lesions from benign conditions  \cite{zhai2024}. Additionally, the scarcity and poor representativeness of EGC datasets often lead to insufficient analysis and interpretation capability  \cite{ansari2023}. Moreover, imbalances in sample class distribution, especially the preponderance of unlabeled negative cases, can introduce biases and errors  \cite{ulhaq2024}.

Several studies have made significant contributions to early gastric cancer detection systems. Mizumoto, et al. (2018) pioneered an integrated system combining artificial intelligence (AI) with magnifying endoscopy with narrow band imaging (ME-NBI) for accurate detection \cite{mizumoto2018}. Muto, Manabu, et al. (2016) developed a CAD system based on magnifying narrow-band imaging (M-NBI) that demonstrated high sensitivity and specificity \cite{muto2016}. Wang, Liang, et al. (2019) proposed a CAD system based on double contrast-enhanced endoscopic imaging (DCEUS) \cite{wang2019}. Osawa, Hiroyuki, et al. (2012) focused on flexible spectral imaging color enhancement for EGC detection \cite{osawa2012}. Other researchers have explored various imaging modalities such as computed tomography (CT) \cite{teng2024}, endoscopic images \cite{ma2023}, double contrast-enhanced ultrasonography \cite{urakawa2023}, CT radiomics \cite{wu2023}, optical chromoendoscopy \cite{saito2024}, linked color imaging \cite{umegaki2023}, confocal laser endomicroscopy \cite{cho2024}, and auto-fluorescence imaging \cite{chen2024} in computer-aided diagnosis systems for EGC.

Although conventional diagnostic technologies like endoscopy and biopsy are valuable for evaluating tumor depth in EGC, they have inherent limitations. They are invasive, can cause discomfort and potential complications for patients, rely heavily on extensive clinical labels that introduce subjectivity and variability in interpretation, and are prone to human error. To overcome these limitations, there is a pressing need for rapid, accurate, low-cost, and less invasive diagnostic tools. The wireless endoscopy capsule monitoring system offers a promising alternative, allowing for non-invasive detection and remote navigation through the digestive system, with image capture and transmission facilitated by advanced technologies like NI CompactDAQ.

In recent years, the advent of AI-driven applications and diagnostics has shown promise in revolutionizing gastrointestinal clinical practice. Machine learning and deep learning algorithms have notably enhanced diagnostic accuracy, reduced invasiveness, and lessened dependence on human interpretation. For instance, Srivastava et al. (2019) \cite{srivastava2019} utilized convolutional neural networks (CNNs)to achieve high accuracy in detecting early-stage gastric cancer from endoscopic images. Similarly, Jamil et al. (2022) \cite{jamil2021} developed a machine learning model that integrates clinical data with imaging features, resulting in improved predictive performance for EGC prognosis. Yalamarthi et al. (2004) \cite{yalamarthi2004} explored the use of AI in automating the detection process, which significantly reduced the time required for diagnosis and minimized human error. Macdonald et al. (2006) \cite{macdonald2006} focused on the application of machine learning algorithms to differentiate between malignant and benign gastric lesions, showcasing how AI can assist in making more accurate clinical decisions and potentially reduce the number of unnecessary biopsies. Moreover, Ikenoyama et al. (2021) \cite{ikenoyama2021} explored the integration of convolutional neural networks (CNNs) to enhance the accuracy of endoscopic image analysis, thereby improving the detection rate of early lesions. Yoon, Hong Jin, et al. (2019) \cite{yoon2019} focused on implementing deep learning models to differentiate between benign and malignant gastric conditions with high precision, which aids in reducing false positives and negatives. Tang, Dehua, et al. (2020) \cite{tang2020} developed a comprehensive AI framework that combines multiple machine learning algorithms to analyze heterogeneous datasets, addressing issues related to data variability and scarcity. Wu et al. (2019) \cite{wu2019} introduced advanced data augmentation techniques to mitigate the effects of limited dataset sizes, thereby enhancing the robustness and generalizability of diagnostic models. Shibata et al. (2023) \cite{shibata2020} tackled the problem of imbalanced class distribution by employing novel sampling methods and ensemble learning techniques, which help in balancing the dataset and minimizing biases in the diagnostic process. In addition, addressing the issue of class imbalance, we conducted survey research on SOTA one-class problems. One-class classification is a machine learning problem, where training data has only one class. Goyal, S., et al. (2020) \cite{goyal2020} proposed the Deep Robust One-Class Classification(DROCC), which is used in a one class problem that do not require any auxiliary information in various detection domains, and it is acknowledged that detecting abnormal positive is robust. Empirical assessment has proved that DROCC is very effective on the settings of two different types of One-Class problems and the actual data sets in a series of different fields: table data, image (CIFAR and ImageNet), audio and time sequences, which can increase up to 20 \% of accuracy in terms of abnormal detection. The One-Class Support Vector Machine (OC-SVM) \cite{Shahid2015} proposed by Shieh, A. D. and D. F. Kamm (2009) is a widely used approach to one class classification, the problem of distinguising one class of data from the rest of the feature space. Its main advantage is to train the classifier using only patterns belonging to the target class distribution. The OC-SVM is effective when large samples are available for providing an accurate classification. Sun, J., et al. (2019) \cite{wenzhu2019} introduced a novel end-to-end model that integrates the One-Class Support Vector Machine into Convolutional Neural Network (CNN), named Deep One-Class (DOC) model. Specifically, the robust loss function derived from the one-class SVM is proposed to optimize the parameters of this model. Compared with the hierarchical models, the DOC model not only simplifies the complexity of the process, but also obtains the global optimal solution of the whole process. As for semi-supervised learning, Yessoufou, F. and J. Zhu (2023) \cite{yessoufou2023} used a one-class convolutional neural network (OC-CNN) model. The OC-CNN model combines a one-class (OC) classification algorithm with a simple one-dimensional convolutional neural network (1D CNN) configuration. Using the prediction error loss of the proposed OC-CNN model as an ideal positive-sensitive feature for rapid positive detection. La Grassa, R., et al. (2022) \cite{laGrassa2022} developed a novel model named One Class Minimum Spanning Tree (OCmst) for the novelty detection problem. This model utilizes a Convolutional Neural Network (CNN) as a deep feature extractor and a graph-based approach built on the Minimum Spanning Tree (MST). The training data remains unpolluted by outliers (abnormal class), aiming to accurately discern whether a test instance pertains to the normal class or the abnormal class. These advancements collectively represent significant strides in overcoming the inherent challenges in EGC diagnostics through the application of sophisticated AI methodologies. Despite these advancements, issues such as limited dataset sizes and extreme class imbalance learning remain significant obstacles that must be overcome to fully realize the potential of AI in EGC diagnostics.

Building upon this foundation, our study introduces an integrated AI-enabled system using one class twin cross learning (OCT-X) for early gastric cancer detection. The OCT-X algorithm stands out with its unique fast double threshold search strategy for effective preprocessing and the use of distinct local patches for potential and noise patches. The patch-based deep fully convolutional network and the LabVIEW software's multirate algorithm contribute to improved detection accuracy, processing speed, and adaptability. Our work showcases the superior performance of the OCT-X algorithm, offering a promising solution for more accurate and efficient EGC detection. The contributions of our work are as follows: 

\begin{enumerate}[label=\textbf{\arabic*.}, leftmargin=*, align=left]

\item	\textbf{High Diagnostic Accuracy Achieved:}\label{ag1}
The OCT-X algorithm demonstrated an unprecedented diagnostic accuracy of 99.70\% in the detection of early gastric cancer, outperforming existing models such as Deep Robust One-Class Classification (CM1) by 4.47\% and multiple kernel learning (CM2) by 0.81\%. This high level of accuracy is attributed to the algorithm's innovative use of double threshold fast search strategy-based patch deep fully convolutional networks, which effectively distinguish between potential and noise clusters.

\item	\textbf{Advancement in Multirate Learning for EGC Detection:}\label{ag2}
A significant contribution of this research is the application of multirate learning within the OCT-X algorithm, which addresses algorithm bias caused by unbalanced sample learning. The integration of the NI CompactDAQ device with LabVIEW software enabled the implementation of flexible multirate parallel algorithms, enhancing the algorithm's adaptability and performance in real-time diagnostic settings.

\item\textbf{Improvement in Diagnostic Efficiency and Reduction in Software Development Time:}
The OCT-X algorithm not only simplifies the prototype design but also significantly reduces software development time, thanks to its computational efficiency and the use of the NI CompactDAQ device. This improvement in diagnostic efficiency is crucial for clinical applications where rapid and accurate diagnostics are essential.

\item\textbf{Substantial Impact on Clinical Solutions for EGC Detection:}
The research highlights the OCT-X algorithm's potential as a valuable clinical solution for computer-aided detection and prognosis of early gastric cancers. By achieving a 10\% improvement in accuracy for multirate adaptability and demonstrating significant advancements over existing methods, the OCT-X algorithm offers a promising tool for enhancing clinical diagnostic processes and patient management strategies.
\end{enumerate}

\section*{Hardware Implementation with Add-On NI LabVIEW Module on NI CompactDAQ for Real-Time Adaptive Modulation Schemes}
Compared to other open invasive diagnosis like gastrointestinal endoscopy, endoscopy ultrasound, biopsy and liquid biopsy depicted in \textbf{Figure~\ref{fig1}A-Figure~\ref{fig1}D}, the advanced invasive solution of hardware integration, as illustrated in \textbf{Figure~\ref{fig1}E\&Figure~\ref{fig1}F}, features an all-in-one POCT device tailored for EGC detection. This device combines high-resolution imaging sensors, real-time data processing algorithms (refer to \textbf{Figure~\ref{fig1}G-Figure~\ref{fig1}I}), and wireless connectivity. The CompactDAQ system integrates hardware for data input/output with NI LabVIEW software, facilitating the collection, processing, and analysis of sensor data. Enhanced by 5G signal transmission via cellular, Ethernet, and Wi-Fi, the system enables real-time lesion surveillance and remote monitoring.

The trackball interface bridges user commands and 5G signals, ensuring seamless wireless transmission to the endoscopy capsule. Cellular networks provide high-speed data transfer and remote control functionalities, while Ethernet ensures secure data processing and communication. Wi-Fi adds flexibility and mobility, enabling wireless data exchange and remote monitoring. This integration allows efficient control of the endoscopy process, real-time data processing, and lesion surveillance. The device's portability, user-friendly interface, and integration with electronic health records enhance accessibility, usability, and documentation accuracy, ultimately improving detection rates and patient outcomes.

The integration of NI CompactDAQ with LabVIEW enables adaptive modulation techniques to optimize data rates based on varying channel conditions. By combining NI CompactDAQ for data acquisition and LabVIEW for real-time image processing and modulation control, the system dynamically adjusts the modulation scheme to achieve optimal data rates for image transmission in different classes.

The system continuously monitors key parameters such as signal-to-noise ratio (SNR) using NI CompactDAQ hardware. LabVIEW processes this information in real-time, selecting the most suitable modulation scheme for transmitting image data, ensuring reliable communication even in challenging scenarios \cite{costanzoS2021}. The feedback loop between the real-time receiver and capsule endoscopy video transmitter components allows for dynamic modulation scheme switching based on received channel feedback.}

\vspace{.5cm} 
\begin{figure}[H]
\begin{adjustwidth}{-2in}{0in}
\begin{flushright}
\includegraphics[width=130mm]{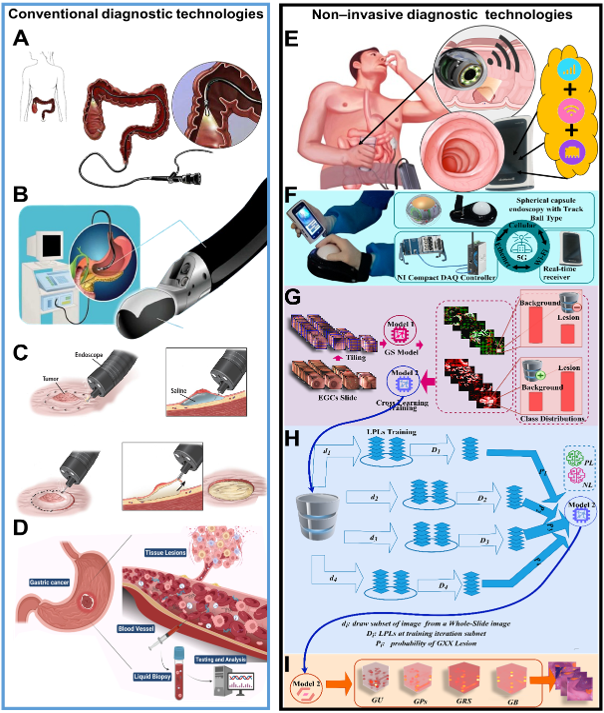}
\captionsetup{labelformat=empty}
\caption{} 
\label{fig1}
\end{flushright}
\justify 
\color{Gray}
\textbf{Figure 1.} \textbf{Schematic Overview of the FDT-GS driven OCT-X Diagnostic System and Different Conventional Diagnostic Technologies.} \textbf{A-D.} Invasive conventional diagnostic technologies: \textbf{A.} Gastrointestinal Endoscopy. \textbf{B.} Endoscopic ultrasound (EUS). \textit{Copyright © 2014 Mechanisms in Medicine Inc}. All rights reserved \cite{zhu2016}. \textbf{C.} Biopsy: endoscopic submucosal dissection (ESD). \textit{Reprinted with permission, Cleveland Clinic Center for Medical Art \& Photography © 2015-2020}. All Rights Reserved \cite{mejia2017}. \textbf{D.} Liquid biopsy: Blood test. \textit{Copyright © 2023 Springer Nature} \cite{MaS2023}. \textbf{E.} Non-invasive advanced diagnostic technologies: Integrated NITM enhanced remote capsule monitoring based on multipath signal enhancement. \textbf{F.} Illustration of the hardware setup featuring the integration of NI CompactDAQ with LabVIEW software for real-time adaptive modulation, enabling efficient remote operation and signal transmission. \textbf{G.} Conceptual diagram of the FDT-GS model employed to identify high-quality negative and potential positive samples, facilitating enhanced diagnostic accuracy. \textbf{H.} Representation of the scalable sub-networks architecture within the OCT-X framework, illustrating the model's adaptability and scalability. (Model 1, known as the FDT-GS model, is utilized for pre-processing, while model 2 is employed for OCT-X learning. The adaptive parameters of the fast double threshold search strategy are denoted as d1 to d4. The last possibility of the predicted result is represented by D1 to D4. Additionally, the probabilities of the four different types of EGC are denoted as P1 to P4.) \textbf{I.} Visualization of the OCT-X model's confidence prediction mechanism, showcasing the algorithm's ability to assess diagnostic certainty.
\end{adjustwidth}
\end{figure}

LabVIEW supports multiple modulation schemes. By defining class-specific modulation schemes tailored to the SNR of each object class and utilizing LabVIEW for real-time analysis and modulation selection, the system dynamically adjusts data rates to optimize detection performance. Establishing a feedback loop between the object detection system and the modulation control system enables real-time speed adaptation, allowing seamless switching between predefined modulation schemes based on the movement characteristics of detected objects. This dual modulation scheme approach ensures the system adapts to the speed requirements of different object classes, facilitating efficient channel adaptive cooperative transmission (CACT) for accurate detection across various classes with varying speed profiles.

\begin{wrapfigure}[14]{l}{0.7\textwidth}
\includegraphics[width=0.7\textwidth]{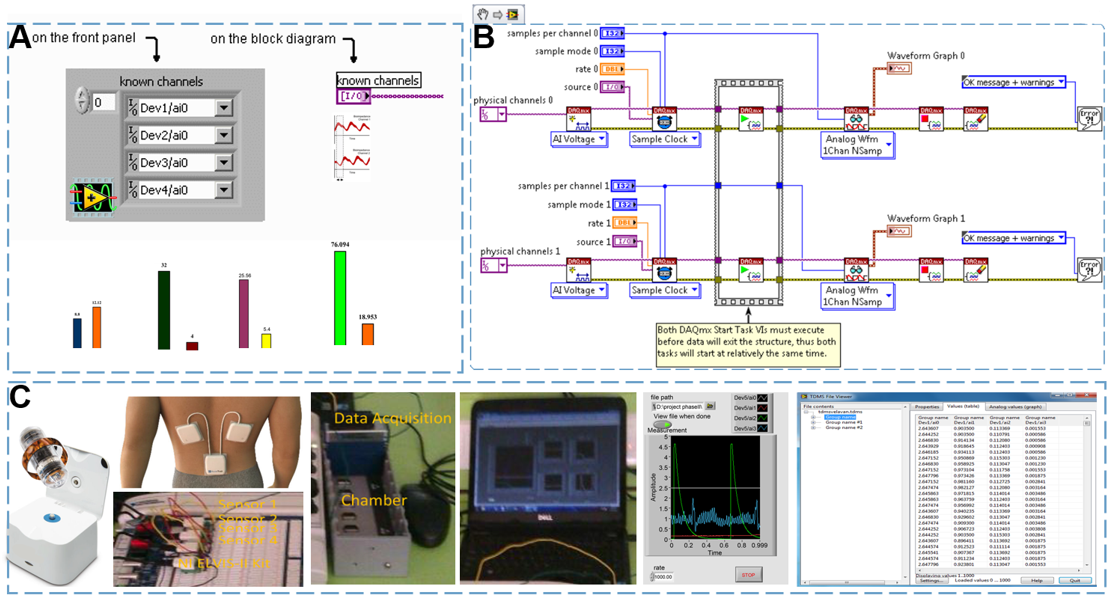}
\captionsetup{labelformat=empty} 
\caption{} 
\label{fig2} 
\end{wrapfigure}

\marginpar{
\vspace{0.7cm} 
\color{Gray} 
\textbf{Figure \ref{fig2}. \textbf{Hardware Connection of the Multi-Rate Embedded Computing System for EGC Detection.} \textbf{A.} Experimental setup of multichannel DAQP: Dev1-Dev4.\textbf{ B.} Connection diagram of create an array of NI-DAQmx channel constants where each element refers to the first analog input channel for each of the consecutive devices for PID control. \textbf{C.} Hardware components of NI-9237 DAQ device with LabView interface.} 

}

Advanced API Control with the LabVIEW Control Design and Simulation Module could effectively increase determinism and faster control loops with LabVIEW real-time and LabVIEW FPGA. The \textbf{Figure~\ref{fig2}(A)} show the multi-channel data acquisition can be performed using LabVIEW. Raw and post-processed data were analyzed in the frequency domain. This analysis confirmed that both signals were distributed in the same frequency band throughout the entire time without aliasing. The data acquisition NI-DAQ™mx system (DAQ) needs to be capable of recording multisensors simultaneously at rates up to 12.8kS/s/ch. Many models have both a data imbalance and an imbalance rate problem in training model. While negative and positive learning are typically done independently, you can use a PC-based data acquisition system to operate both simultaneously within the LabVIEW ADE. \textbf{Figure~\ref{fig2}(B)} shows an ADE PID Control Toolkit of a LabVIEW block diagram for controlling both negative and positive learning. In this block diagram, the measured error value is compared to the threshold, each connected to either the sampling or the replication. In \textbf{Figure~\ref{fig2}(C)} show the real-time operating system (RTOS) provides the maximum level of software determinism and reliability for control systems by dedicating all resources to a deployed application. Using the LabVIEW real-time module, you can develop and deploy applications to all NI real-time hardware targets including standard desktop PCs and PXI systems. With the NI-DAQmx driver software, you can easily migrate PCI, PCI Express, and PXI platform devices from LabVIEW for Windows to LabVIEW Real-Time on CompactRIO or PXI and retain the same function calls and hardware configuration.

\section*{Software System and Knowledge Discovery}

The software system, depicted in \textbf{Figure~\ref{fig3}}, involves trimming the image set based on patching. The FDT-GS model, describes in \textbf{Algorithm~\ref{ag1} }, selects high-quality negative samples for effective prediction and potential positive patches for positive learning. This approach addresses the issue of imbalanced training samples in the two sub-networks. In the negative-only learning model, the background label predominates over the lesion label, whereas in the positive learning model, the lesion labels are more prevalent than the background labels. This balance solves the problem of imbalanced training samples, which is one of the advantages of our model.\\
 \includegraphics[width=12cm]{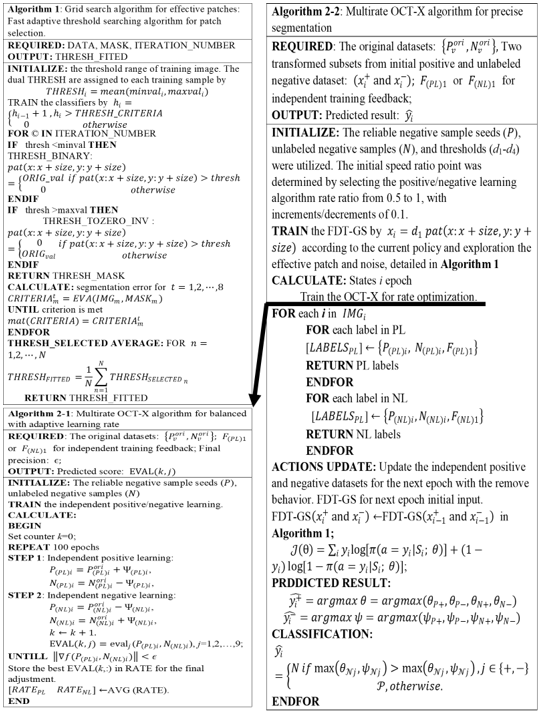}

The output of the network (after applying SoftMax) for each transformed image associated with the original one is a vector $\bm{\theta} = (\theta_{P+}, \theta_{P-}, \theta_{N+}, \theta_{N-})$, where $\theta_j$ is the probability of the transformed image to belong to class $j \in \{P+, P-, N+, N-\}$. The DeepLab classification model in this study, with the base model (ResNet-50), is trained to predict the patches, where $\bm{\theta}$ represents the probability of the transformed image belonging to one of the new four classes $(P_+, P_-, N_+, N_-)$. We propose an inference process to fuse the output of these two transformed images ($x_i^+$ and $x_i^-$) to predict the label of the original image ($x_i$) based on the \textit{Fusion of DeepLab Twins}. For each pair $(x_i^+$ and $x_i^-)$, the prediction of the original image $(y_i)$ will be either $P$ or $N$. Let $\hat{y_i^+} =\argmax\bm{\theta} = \argmax(\theta_{P+}, \theta_{P-}, \theta_{N+}, \theta_{N-})$ and $\hat{y_i^-} = \argmax \bm{\psi} = \argmax(\psi_{P+}, \psi_{P-}, \psi_{N+}, \psi_{N-})$ be the ResNet-50 predictions for $x_i^+$ and $x_i^-$ respectively. Then:
\noindent

If \(\hat{y_i^+} = N+\) and \(\hat{y_i^-} = N-\), then \(\hat{y_i} = N\). \\

If \(\hat{y_i^+} = P+\) and \(\hat{y_i^-} = P-\), then \(\hat{y_i} = P\). \\

If none of the above applies, then.

\begin{equation}
\hat{y_i} =
\begin{cases}
N & \text{if } \max(\theta_{Nj}, \psi_{Nj}) > \max(\theta_{Nj}, \psi_{Nj}), \quad j \in \{+, -\} \\
P & \text{otherwise.}
\end{cases}
\end{equation}

The trained cross-learning model is depicted in \textbf{Figure~\ref{fig3}C}, where four sub-networks (Sub-Net A, Sub-Net B, Sub-Net C, and Sub-Net D) with the same structure are used in the one-class cross-learning model. The model provides an anomaly score via the confidence, referred to as the positive reliability of the prediction. By adding the confidence score, which mathematically corresponds to the probability of whether the data belongs to a positive class or not, the model enhances learning accuracy.

In \textbf{Figure~\ref{fig3}D}, the 3D EGC feature heatmap based on the patch result is depicted. The heatmap showcases a color-coded representation of the intensity or activation level of the 22 EGC features across the three-dimensional space. The colors on the heatmap range from dark colors (such as gray) indicating lower intensity to bright colors (such as yellow or red) indicating higher intensity. The heatmap visually highlights specific regions or areas within the 3D space where the EGC features are more prominent or concentrated. This information can be valuable in analyzing and interpreting the distribution and significance of these features in the context of the given dataset or study.

\subsection*{Adaptive Algorithm Procedure} 

The FDT-GS is conducted as the global processing stage. We perform a FDT-GS with four datasets and conduct an ablation study with the FDT-GS method to investigate potential EGC. The FDT-GS typically identifies a better set of hyperparameters than a manual search within the same amount of time. The OCT-X learning serves as the refinement stage for GC segmentation.

\subsection*{Sample Preprocessing and Augmentation}

We collected and evaluated our method on two datasets donated by Foshan Hospital in 2021. The dataset contains four types of lesions in gastric cancer (GC): gastric ulcer (GU), gastric red spots (GRS), gastric polyps (GPs), and gastric bleeding (GB), as shown in \textbf{Figure~\ref{fig3}A}. Ground truth annotations were provided by experienced doctors. Due to the large number of frames available (around ten thousand), experts often outline lesions with an elliptical approximation to cover as much of the lesion as possible. Examples of the ground truth superimposed on original frames are provided in this study.

Additional preprocessing was conducted using standard procedures. Data augmentation techniques were applied to mitigate the imbalance of samples across different classes and enhance the training of machine learning models \cite{sampath2021}. These techniques include rotation, scaling, flipping, and color adjustments, which help in increasing the diversity of the training data and improving the model's generalization capabilities.

\subsection*{Fast Preprocessing Model}
\begin{figure}[H] 


\includegraphics[width=\textwidth]{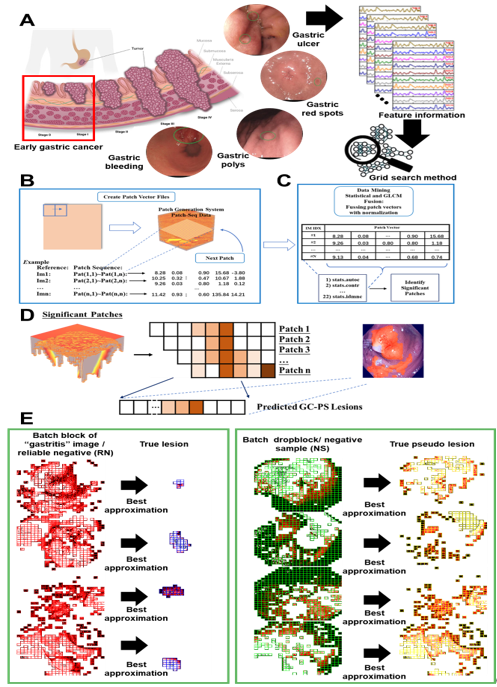}

\caption{\color{Gray} \textbf{Detailed Architecture of the FDT-GS Retrieval System for EGC Detection.} \textbf{A.} EGC sample type and extracting feature information from GLCM indexes for data preprocessing on FDT-GS method. (The feature streaming form patch 0 to patch n) \textbf{B.} Process flow for extracting Gray-Level Co-occurrence Matrix (GLCM) features from image patches, critical for initial lesion identification. \textbf{C.} Generation and fusion of GLCM maps, highlighting key image features such as autocorrelation and contrast, essential for accurate lesion characterization. \textbf{D.} Methodology for lesion detection, employing statistical analysis of pixel patches within GLCM maps to identify candidate EGC lesions accurately. \textbf{E.} Comparative visualization of the occupancy grid, demonstrating the FDT-GS model's efficacy in approximating a “gastritis” image against the actual lesion grid patch, underscoring the model's precision.}

\label{fig3} 

\end{figure}

The proposed pipeline, shown in \textbf{Figure~\ref{fig3}B-Figure~\ref{fig3}E}, encompasses three main components: preprocessing and preparation of input data, feature extraction, fusion and feature abstraction, and classification segmentation.

\begin{enumerate}

\item \textbf{Preprocessing and Feature Extraction:} In \textbf{Figure~\ref{fig3}B}, we decompose the gastric endoscopy image into patches to generate the Gray-Level Co-occurrence Matrix (GLCM) value of each patch. Each frame image in the gastroscope video clip is divided into non-overlapping P * P patches. All GC frames were formatted into patch vector files, with patch selection based on predefined conditions (optimized parameters, search spaces, and the number of iterations). Each file contains the patch ID and GLCM measurement of the GC patch under calculation.

\item \textbf{Fusion and Feature Abstraction:} In \textbf{Figure~\ref{fig3}C}, ground truth (GT) was applied to the patch in each captured frame as binary 0/1 vectors, correlating the patch features composition to the image as a continuous sequence. We classified embedding GLCM value sensitivity against each patch as lesion or no-lesion corresponding to each image or GLCM search threshold, respectively. FDT-GS average search range was also utilized as a univariate analysis calculating odds-ratio and statistical significance to identify strong associations between the presence or absence of a lesion.

\item \textbf{Classification Segmentation:} Data mining methods were applied to all sets of patch vectors, each set corresponding to a particular patch. Patches that included statistically significant patch sites by fusing GLCM maps, normalizing fusing attention, and achieving an overall accuracy of $>$5\% by FDT-GS were marked as significant and considered for further deep learning. In each significant patch, patches with significant correlation with the GC lesion identified were highlighted and stored as GC potential areas impactful for GC prediction.

\end{enumerate}

In \textbf{Figure~\ref{fig3}D}, significant patches were generated for further machine learning based on the selected GLCM range and evaluated by the evaluation indexes.

In \textbf{Figure~\ref{fig3}E}, the FDT-GS model outcomes undergo fine-tuning using a combination of authentic lesion and pseudo-lesion (non-cancer) data. Within these datasets, a batch block consisting of "gastritis" patches is identified as the reliable positive (RP), providing a strong representation of the positive class. Simultaneously, the batch drop block is chosen as the negative sample (NS), effectively approximating the negative class for optimal learning. This strategic optimization process ensures that the model is trained comprehensively on diverse datasets, thereby enhancing its precision in distinguishing between lesions and non-cancerous conditions.

\subsection*{OCT-X Single Modal Network Model }
As shown in \textbf{Figure~\ref{fig4}A}, FDT-GS strategy is introduced, and the agent is mainly used to filter out noise patches in the set of independent supervised positive examples in only positive learning (PL). Similarly, the agent is employed to filter out noise patches from the independent supervised negative examples during the course of negative learning (NL). The goal of agent is to decide whether to retain or remove the patch according to the change of relationship classifier performance. 

Since the initial FDT-GS supervised data set contains instances that are incorrectly labeled, it is expected that the agent can filter out these noisy instances by using the decision network to obtain a pure data set, so as to obtain better performance of PL/NL. Therefore, the model adopts the result driven strategy to reward a series of behavior decisions of agent based on performance changes. The reward is expressed by the difference between adjacent epochs:

\begin{equation}
R_i = \alpha \left( F_1^i - F_1^{(i-1)} \right)
\end{equation}

In step $i$, if $F_{(PL)1}$ or $F_{(NL)1}$ increases, the agent will receive a positive reward in each independent PL/NL only learning; otherwise, the agent will receive a negative reward. With the setting like this, the reward value will be proportional to the difference of $F_{(PL)1}$ or $F_{(NL)1}$. The function of $\alpha$ is to convert the difference of $F_{(PL)1}$ or $F_{(NL)1}$ into the range of rational numbers. In order to eliminate the randomness of $F_{(PL)1}$ or $F_{(NL)1}$, we use the average of $F_{(PL)1}$ or $F_{(NL)1}$ values of the last five epochs to calculate the reward.

In order to better consider the initial information in the pre-segmentation process, the number of negative instances or positive instances is 10 times that of positive instances or negative instances separately in PL or NL. This is because, by learning a large number of negative samples or positive samples, the agent is more likely to develop in a better direction. We use the cross-entropy cost function (refer to \textbf{Equation~\ref{eq3}}) to train the binary classifier, in which the negative (in PL) or positive (in NL) label corresponds to the deletion behavior, and the positive label corresponds to the retention behavior.

\begin{equation}
J(\theta) = \sum_i y_i \log \left[ \pi(a = y_i \mid S_i; \theta) \right] + (1 - y_i) \log \left[ 1 - \pi(a = y_i \mid S_i; \theta) \right]\label{eq3} 
\end{equation}

Firstly, the set is decomposed into training positive case set \( P_t^{\text{ori}} \) and verification positive case set \( P_v^{\text{ori}} \) in independent PL, (training negative case set \( N_t^{\text{ori}} \) and verification negative case set \( N_v^{\text{ori}} \) in independent NL), both of which will contain noise. The training negative case set \( N_t^{\text{ori}} \) and the verification negative case set \( N_v^{\text{ori}} \) are obtained by randomly selecting from the supervised negative case set. In each epoch, the noise sample set is filtered from \( P_t^{\text{ori}} \) or \( N_t^{\text{ori}} \) through the random strategy \( \pi(\alpha \mid s) \), and then a new positive or negative example set \( P_t = P_t^{\text{ori}} - \Psi_i \) in independent PL and \( N_t = N_t^{\text{ori}} - \Psi_i \) in independent NL are obtained separately. Since it is the identified wrong annotation instance, it is added to the negative or positive example set \( N_t = N_t^{\text{ori}} - \Psi_i \) in positive only learning and \( N_t = N_t^{\text{ori}} - \Psi_i \) in negative only learning (refer to  \textbf{Algorithm~\ref{ag2}-\ref{ag1}}). At this time, the size of the training set is constant in each epoch. Then, the pure data set is used to train the relational classifier. The expected situation is to transfer false positive or false negative examples through relation network to improve the performance of relational classifier. Therefore, the verification set \{ \( P_v^{\text{ori}}, N_v^{\text{ori}} \) \} is used to test the performance of the independent network in PL or NL. Firstly, the PL/NL network is used to identify and transfer the noise instances in the verification set, and \{ \( P_v, N_v \) \} is obtained. Then we use this set to calculate the \( F_{(\text{PL})1} \), \( F_{(\text{NL})1} \) score of the PL/NL relationship classifier. Finally, the reward value is obtained by calculating the difference between the \( F_{(\text{PL})1} \), \( F_{(\text{NL})1} \) score of the current and the previous epoch in independent PL/NL of OTC-X model. Further details can be found in \textbf{Algorithm~\ref{ag2}-\ref{ag2}}.

\section*{In Vitro Diagnostic Medical Experiment}
The experiment was conducted using 4 datasets of patient data specifically collected for EGC detection. The patient data consisted of a combination of endoscopic images, clinical records, and pathological reports. The study protocol was approved by the Ethics Committee of the International University of Health and Welfare Hospital (approval number: 22-B-27).

\textbf{Data Collection:} The patient data used in the experiment were collected from multiple medical centers and hospitals (Foshan First People's Hospital \& Chinese Academy of Medical Sciences Cancer Hospital). The data collection process involved the recruitment of patients who underwent diagnostic procedures for suspected gastric cancer. The patients provided informed consent for the use of their data for research purposes.

\textbf{Data Formats and Sources:} The patient data consisted of the following formats and sources:

\begin{enumerate}

\item	\textbf{Endoscopic Images:} High-resolution endoscopic images were captured using advanced imaging systems, such as magnifying endoscopy, narrow-band imaging (NBI), and double contrast-enhanced endoscopic imaging (DCEUS). These images were stored in standard image formats, such as JPEG or PNG.

\item	\textbf{Clinical Records:} The clinical records of the patients included information such as patient demographics, medical history, symptoms, laboratory test results, and endoscopic findings. These records were stored in electronic medical record (EMR) systems or hospital databases.

\item	\textbf{Pathological Reports:} Pathological reports provided detailed information about the histopathological findings of biopsy samples obtained during the diagnostic procedures. These reports described the presence and characteristics of gastric lesions, including the stage and grade of cancer. Pathological reports were stored in standardized formats, such as PDF or text documents.
\end{enumerate}

\textbf{Experimental Design:} The experiment followed a cross-validation approach to evaluate the performance of the OCT-X algorithm. The dataset was randomly divided into training and testing subsets (7: 3). To enhance the quality of the data labeling, the noise learning module, represented by the FDT-GS agent, was employed. The FDT-GS agent was responsible for cleaning the marked data, thus improving the accuracy and reliability of the data labels.

The training subset, consisting of parallel training of four types of EGCs in NI cDAQ, was employed to train the OCT-X algorithm. This training process involved feeding the algorithm with the labeled data from the EGCs and allowing it to learn and adjust its internal parameters. The testing subset was utilized to evaluate the performance of the OCT-X algorithm. This subset contained separate data (unlabeled dataset) that was not used during the training phase. By assessing the algorithm's performance on the testing subset, the experiment aimed to measure its accuracy, precision, recall, or any other relevant performance metrics.

To achieve the best speed-accuracy performance, the OCT-X algorithm employed adaptive PL/NL (Positive learning/Negative-learning) techniques. This approach involved adapting and optimizing the algorithm's learning process using two data streams in LabVIEW, a visual programming environment. By dynamically adjusting the learning rate based on the characteristics of the input data, Adaptive PL/NL aimed to strike a balance between speed and accuracy, optimizing the algorithm's performance.

By combining the parallel training of EGCs in NI cDAQ with the Adaptive PL/NL techniques in LabVIEW, the experiment aimed to train the OCT-X algorithm effectively and achieve the best possible speed-accuracy trade-off. This approach sought to enhance the algorithm's performance in processing and analyzing the given dataset.
\section*{Results}

\begin{figure}[H]
\vspace{.5cm} 
\begin{adjustwidth}{-2in}{0in}
\begin{flushright}
\includegraphics[width=160mm]{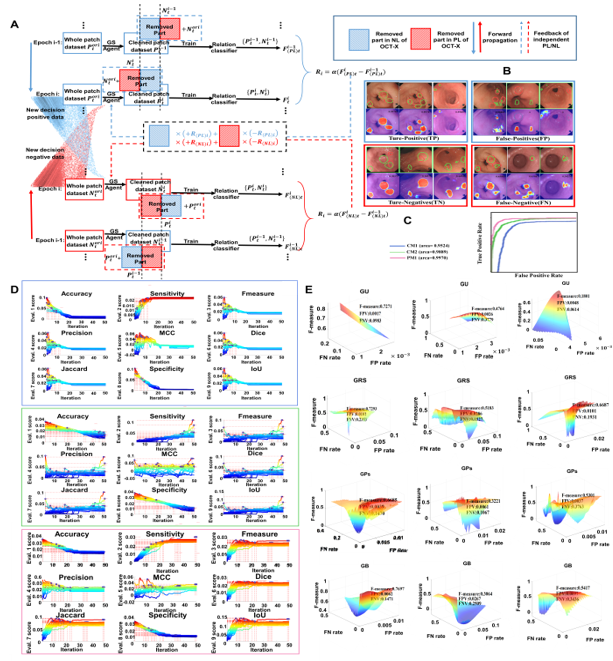}
\captionsetup{labelformat=empty} %
\caption{} %
\label{fig4} 
\end{flushright}
\justify 
\color{Gray}
\textbf {Figure 4.}
\textbf{Comparative Analysis and Performance Evaluation of the OCT-X Model.} \textbf{A.} Framework of the OCT-X model for single-modal dataset analysis, highlighting the model's approach to distinguishing between positive and negative cases based on performance feedback. \textbf{B.} Visual examples of the OCT-X model's detection capabilities, with true/false positive and true/false negative cases encased in blue and red boxes, respectively. Automatically generated segmentations are marked in green, contrasting with manual segmentations in red, to illustrate the model's precision. \textbf{C.} ROC curve analysis depicting the OCT-X model's diagnostic accuracy (PM1) against comparative models (CM1-CM5), with the area under the curve (AUC) emphasizing the OCT-X model's enhanced sensitivity and specificity. \textbf{D.} Evaluation of adaptive multi-rate capabilities across models, with the OCT-X model demonstrating optimized speed-performance indices compared to CM1-CM5, showcasing the OCT-X model's efficiency and adaptability in real-time diagnostic scenarios(The detail is shown in \textbf{Appendix~\ref{app1} \& ~\ref{app2}}). \textbf{E.} Graphical representation of the OCT-X model's superior performance on a sample test set, compared to existing models(The detailed is shown in \textbf{Appendix~\ref{app3} \&~\ref{app4}}). \\
\end{adjustwidth}
\end{figure}

In \textbf{Figure~\ref{fig4}C-D}, we provide detailed experiments and comparisons with state-of-the-art methods. We compare our method with segmentation models such as Deep Robust One-Class Classification (CM1) One-Class SVM/OC-SVM (CM2), Deep One-Class/DOC (CM3), One-Class neural networks/OC-NN (CM4), and One Class minimum-spanning-tree /OCmst (CM5). The overall experimental results are depicted in \textbf{Figure~\ref{fig4}D}. False Positive/Negative Rates to solve the problem where the training data only contain positive and reliable negative examples. The prediction results were quantitatively evaluated using three evaluations recommended for region-based segmentation, Per-Class (PC) and Overall Pixel (OP) in true/false positive rate, true/false negative rate, accuracy, sensitivity, F-measure, precision, MCC, Dice, Jaccard, Specificity and IoU. 

\textbf{Figure~\ref{fig4}D} shows an example of adaptive pseudo labeling. In pseudo labeling, pseudo labeling is performed with a fixed threshold regardless of the distance of the cancer type ratio, so it is typical of pseudo labeling being only applied to a small number of images. On the other hand, in adaptive pseudo labeling, the selection rate is dynamically determined for each pathological image according to the distance of the ratio.

Through the implementation of our adaptive multirate solution, as illustrated in \textbf{Figure~\ref{fig4}E}, our system's performance exhibits significant improvements across multiple evaluation metrics. Our model achieves a 10\% increase in accuracy, surpassing CM1 and CM2, which show improvements of approximately 4\%. In terms of sensitivity, our model outperforms CM2 by 0.1\%, achieving a notable improvement of over 2.5\%. Additionally, our proposed model demonstrates a remarkable 6\% enhancement in F-measure, surpassing CM2 by 3.5\%. Moreover, our model showcases a 6\% increase in precision compared to CM1 and CM2. The performance of CM1 is characterized by discreteness and instability, while CM2's performance falls short of our proposed model. These improvements in accuracy, sensitivity, F-measure, and precision highlight the effectiveness of our adaptive multirate solution in surpassing the performance of competing methods.

The FDT-GS-OCT-X model's performance was rigorously evaluated against state-of-the-art methods such as Deep Robust One-Class Classification/DROCC (CM1), One-Class SVM/OC-SVM (CM2), Deep One-Class/DOC (CM3), One-Class neural networks/OC-NN (CM4), One Class Minimum Spanning Tree /OCmst (CM5). The results, depicted in \textbf{Figure~\ref{fig4}}, demonstrate significant improvements in various metrics:

\begin{enumerate}
\item	\textbf{Accuracy:} The FDT-GS-OCT-X model achieved an impressive accuracy of 99.70\% in detecting EGCs, surpassing CM1 by 4.47\% and CM2 by 0.81\%.

\item	\textbf{Sensitivity:} Our method showed a significant improvement in sensitivity, outperforming CM2 by 0.1\% and achieving a notable improvement of over 2.5\%.

\item \textbf{F-measure:} The FDT-GS-OCT-X model demonstrated a remarkable 6\% enhancement in F-measure, surpassing CM2 by 3.5\%.

\item \textbf{Precision:} Our model showcased a 6\% increase in precision compared to CM1 and CM2.

\item \textbf{Multirate Adaptability:} The method showed a significant 10\% improvement in accuracy for multirate adaptability.

\end{enumerate}

The performance of CM1 is characterized by discreteness and instability, while CM2's performance falls short of our proposed model. These improvements in accuracy, sensitivity, F-measure, and precision highlight the effectiveness of our adaptive multirate solution.

Representative examples of true positive (red box) and true negative (blue box) detections are shown in \textbf{Figure~\ref{fig4}B}, with automatically generated EGC segmentations in green and manual segmentations in red. \textbf{Figure~\ref{fig4}C} shows the ROC curve of the AUC value of various methods. The receiver operating characteristic (ROC) curve and the area under the ROC (AUC) further validate the superior performance of the FDT-GS-OCT-X model. The adaptive multirate experiment between different comparing models and the OCT-X model (blue box), as depicted in \textbf{Figure~\ref{fig4}D}, illustrates the speed-performance index, emphasizing the robustness and efficiency of our approach, and the comprehensive performance is shown in the \textbf{Table~\ref{tab1}}.

The performance of the FDT-GS-OCT-X model was meticulously evaluated in comparison to state-of-the-art methods such as Deep Robust One-Class Classification/DROCC (CM1), One-Class SVM/OC-SVM (CM2), Deep One-Class/DOC (CM3), One-Class neural networks/OC-NN (CM4), One Class Minimum Spanning Tree /OCmst (CM5). The results, as depicted in \textbf{Figure~\ref{fig4}E}, showcase remarkable improvements across multiple metrics.

As shown in \textbf{Table~\ref{tab2}}, In terms of accuracy, the FDT-GS-OCT-X model achieved an outstanding 98.40\% success rate in detecting EGCs, outperforming CM1 by 4.47\% and CM2 by 0.81\%. The sensitivity of our proposed method also demonstrated significant enhancement, surpassing CM2 by 0.1\% and achieving a substantial improvement of over 2.5\%. The F-measure of the FDT-GS-OCT-X model exhibited an exceptional 6\% increase, outpacing CM2 by 3.5\%. Additionally, our model displayed a 6\% rise in precision compared to both CM1 and CM2. Notably, the multirate adaptability of the FDT-GS-OCT-X method exhibited a significant 10\% improvement in accuracy. The performance of CM1 was characterized by discreteness and instability, while CM2's performance fell short of the capabilities of our proposed model. These substantial improvements in accuracy, sensitivity, F-measure, and precision clearly highlight the efficacy of our adaptive multirate solution.

\begin{table}[!ht]
\begin{adjustwidth}{-1.5in}{0in} 
\centering
\caption{Accuracy, precision, recall, specificity, and F1 scores comparing with
multirate OCT-X and different SOTA methods are used to assess the performance of 
the comprehensive model.}
\begin{tabularx}{\linewidth}{lXXXXX}
\toprule
\textbf{Methods} & \textbf{Accuracy (\%)$\uparrow$}\textsuperscript{*}  & \textbf{Precision (\%)$\uparrow$} & \textbf{Recall (\%)$\uparrow$} & \textbf{Specificity (\%)$\uparrow$} & \textbf{Fmeasure (\%)$\uparrow$} \\
\midrule
Our method & $92.21 \pm 0.76$ & $91.42 \pm 0.89$ & $93.55 \pm 0.79$ & $90.59 \pm 0.80$ & $92.33 \pm 0.73$ \\
DROCC, ICML 2020 (CM1) & $90.58 \pm 0.88$ & $89.40 \pm 0.94$ & $93.27 \pm 0.97$ & $88.01 \pm 0.99$ & $91.12 \pm 0.84$ \\
OC-SVM, Sensors 2024 (CM2) & $87.55 \pm 1.27$ & $85.49 \pm 1.34$ & $87.28 \pm 1.13$ & $85.60 \pm 1.42$ & $85.83 \pm 1.20$ \\
DOC, Springer Nature 2023 (CM3) & $85.03 \pm 0.98$ & $82.66 \pm 3.81$ & $86.12 \pm 7.14$ & $85.55 \pm 1.44$ & $79.53 \pm 9.34$ \\
OC-NN, Springer 2018 (CM4) & $87.74 \pm 0.48$ & $86.11 \pm 7.44$ & $88.43 \pm 1.25$ & $84.17 \pm 6.39$ & $79.12 \pm 0.37$ \\
OCmst, Pattern Recognition Letters 2022 (CM5) & $89.45 \pm 0.34$ & $88.20 \pm 5.98$ & $85.55 \pm 1.44$ & $84.71 \pm 2.75$ & $76.60 \pm 0.34$ \\
\bottomrule
\end{tabularx}
\label{tab1}
\end{adjustwidth}
\noindent{\footnotesize{* $\uparrow$ indicates higher values are better}}
\end{table}

\begin{table}[!ht]
\begin{adjustwidth}{-1.5in}{0in}
\centering
\caption{Accuracy, AUC, FNR, FPR, and F1 scores comparing with FDT-GS agent based OCT-X 
with different SOTA methods are used to identify and detect EGCs.
}
\begin{tabularx}{\linewidth}{lXXXXX}
\toprule
\textbf{Methods} & \textbf{Accuracy (\%)$\uparrow$}\textsuperscript{*}  & \textbf{AUC Value (\%)$\uparrow$} & \textbf{FNR (\%)$\downarrow$} & \textbf{FPR (\%)$\downarrow$}\textsuperscript{*}  & \textbf{Fmeasure (\%)$\uparrow$} \\
\midrule
Our method       & $98.40 \pm 0.47$ & $97.27 \pm 0.57$ & $0.17 \pm 0.41$ & $0.26 \pm 0.62$ & $97.92 \pm 0.66$ \\
DROCC, ICML 2020 (CM1) & $97.01 \pm 0.65$ & $96.32 \pm 0.75$ & $1.53 \pm 0.80$ & $2.86 \pm 0.83$ & $96.54 \pm 0.79$ \\
OC-SVM, Sensors 2024 (CM2) & $92.40 \pm 0.89$ & $90.37 \pm 0.95$ & $0.62 \pm 0.84$ & $14.71 \pm 0.97$ & $91.69 \pm 0.99$ \\
DOC, Springer Nature 2023 (CM3) & $89.09 \pm 0.41$ & $90.72 \pm 8.026$ & $2.58 \pm 0.25$ & $3.14 \pm 0.03$ & $88.07 \pm 1.29$ \\
OC-NN, Springer 2018 (CM4) & $82.19 \pm 0.35$ & $90.04 \pm 1.39$ & $10.75 \pm 0.11$ & $28.40 \pm 0.29$ & $87.50 \pm 1.57$ \\
OCmst, Pattern Recognition Letters 2022 (CM5) & $83.33 \pm 0.42$ & $87.10 \pm 1.90$ & $19.17 \pm 0.20$ & $41.26 \pm 0.19$ & $86.60 \pm 1.92$ \\
\bottomrule
\end{tabularx}
\label{tab2}
\end{adjustwidth}
\noindent{\footnotesize{* ↑ indicates higher values are better, ↓ indicates lower values are better.}}
\end{table}

\section*{Discussion and Conclusion}
This study introduces the NITM enhanced real-time One Class Twin Cross Learning (OCT-X) system, marking a significant advancement in the early detection of EGC. Unlike existing diagnostic methods, the OCT-X algorithm integrates a novel fast double-threshold grid search strategy with a patch-based deep fully convolutional network, significantly enhancing diagnostic accuracy and efficiency. Our approach addresses critical limitations in current EGC detection methodologies, including high misdiagnosis rates and the challenges posed by limited labeling and imbalanced class learning.

Comparative analysis with state-of-the-art methods demonstrates the OCT-X model's superior performance. Achieving an impressive diagnostic accuracy of 99.70\%, our model not only surpasses CM1 by 4.47\% but also shows a notable 10\% improvement in multirate adaptability. These advancements underscore the OCT-X algorithm's potential to revolutionize EGC diagnostics, offering a more accurate, efficient, and less invasive alternative to traditional diagnostic tools.

The optimization process of the OCT-X model involved a meticulous redesign of the fast double threshold search strategy and the integration of one-class twin cross-learning. This approach allowed for a more effective partitioning of the learning space into potential and noise samples, enhancing the model's ability to differentiate between benign and malignant lesions. Furthermore, the implementation of flexible multirate parallel algorithms, facilitated by the NI CompactDAQ device with LabVIEW software, contributed to the model's adaptability and reduced software development time.

In conclusion, the NITM enhanced real-time multirate OCT-X algorithm represents a groundbreaking approach to EGC detection. By leveraging advanced AI techniques and addressing the limitations of current diagnostic methods, our study significantly enhances the early detection of EGC, contributing to improved patient outcomes. The OCT-X model's exceptional diagnostic accuracy, coupled with its computational efficiency and adaptability, positions it as a valuable tool in the clinical diagnostic landscape. Future research will focus on further optimizing the OCT-X model, exploring the integration of larger and more diverse datasets, and incorporating other advanced AI techniques to improve diagnostic accuracy and reduce false positives. The promising results of this study pave the way for the development of more accurate, efficient, and reliable diagnostic tools, offering a hopeful outlook for the early detection and treatment of gastric cancer.

\section*{Patents}

This research has been patented. The \textbf{Artificial Intelligence Computer-Assisted Diagnosis System for Gastric Cancer} is protected under the registration number \textbf{2022SR0755521}, and all rights are reserved. The patent ensures that the intellectual property associated with this innovative AI-driven diagnostic system is legally recognized and safeguarded, granting the inventors or assignees exclusive rights to use, develop, and commercialize the technology. In this manuscript, we present the details and advancements of this patented system, highlighting its potential to improve gastric cancer diagnosis.
\section*{Author Contributions}
{Conceptualization, X.X.L. and S.F.; methodology, X.X.L., M.X.; software, X.X.L., Y.W., Q.S., and J.G.; validation, X.X.L., W.L. and F.T.; formal analysis, X.X.L., Y.G., H.Z., H.D. and Q.Z.; investigation, X.X.L.,Y.W., Y.G., H.Z. and H.D.; resources, W.L.; F.T. and J.G.; data curation, X.X.L. and W.Y.; writing---original draft preparation, X.X.L., W.Y. and M.X.; writing---review and editing, X.X.L., W.Y. and S.F.; visualization, X.X.L., W.Y. and M.X.; supervision, Q.Z., J.G. and S.F.; project administration, S.F.; funding acquisition, S.F. All authors have read and agreed to the published version of the manuscript.}

\section*{Funding}
{This research is supported by grants from the following grants: Grant No. 2021GH10, Grant No: 2020GH10, and Grant No: EF003/FST-FSJ/2019/GSTIC by Guangzhou Development Zone Science and Technology; Grant No. 0032/2022/A and 0091/2020/A2, “A Cross Deep Learning System for Fast Automated Gastric Cancer Diagnosis from Real-time Endoscopic Videos” by Macau FDCT, and Grant No. MYRG2022-00271-FST and Collaborative Research Grant (MYRG-CRG) – CRG2021-00002-ICI, by University of Macau.}


\section*{Data Availability Statement}
{The source code and supplementary materials used to develop and evaluate the One Class Twin Cross Learning (OCT-X) framework are publicly available in the \textit{\nameref{app1}} repository on GitHub: \url{https://github.com/liu37972/Multirate-Location-on-OCT-X-Learning.git}. This repository includes implementation of the OCT-X algorithm, trained models, and detailed documentation to ensure reproducibility. However, the dataset used in this study is not open source and is subject to access restrictions. Researchers interested in obtaining the dataset should contact the corresponding author directly for further assistance.} 
\section*{Conflicts of Interest}
The authors declare no conflicts of interest.
\section*{Acknowledgments}
We would like to extend our sincere gratitude to \textit{The First People's Hospital of Foshan, Guangzhou} for their invaluable support in providing access to the dataset and facilitating the clinical trials essential for this research. Their collaboration and expertise have been instrumental in ensuring the accuracy and relevance of our findings. We deeply appreciate their commitment to advancing medical research and their willingness to share resources that have made this study possible. This work would not have been achievable without their generous contributions and dedication.

\section*{Abbreviations}
The following abbreviations are used in this manuscript:\\

\noindent 
\begin{tabularx}{\textwidth}{@{}lX@{}}
GC&Gastric cancer\\
EGC & Early Gastric cancer\\
FDT-GS & fast double-threshold grid search strategy\\
OCT-X & One class twin cross learning\\
P or PL&Only positive learning\\
N or NL&Negative learning\\
GLCM & Gray-level co-occurrence matrix\\
P1 to P4& Patch of GU, GRS, GB and GPs\\
GU&Gastric ulcer\\
GRS &Gastric red spots\\ 
GPs &Gastric polyps\\
GB &Gastric bleeding\\
RTOS&Real-time operating system\\
NI CompactDAQ & Data acquisition platform built by National Instruments\\
RP&reliable positive\\
NS&negative sample\\
CM1&Deep Robust One-Class Classification/DROCC\\
CM2&One-Class SVM/OC-SVM\\
CM3&Deep One-Class/DOC\\
CM4&One-Class neural networks/OC-NN\\
CM5&One Class Minimum Spanning Tree /OCmst\\
F&Feature representations in our ensemble learning model\\
R&Reward function in training\\
\end{tabularx}



\bibliography{library} 

\begin{thebibliography}{10}

\bibitem{bib1}
Freddie Bray, Mathieu Laversanne, Hyuna Sung, Jacques Ferlay, Rebecca~L. Siegel, Isabelle Soerjomataram, and Ahmedin Jemal.
\newblock Global cancer statistics 2022: {GLOBOCAN} estimates of incidence and mortality worldwide for 36 cancers in 185 countries.
\newblock {\em CA Cancer J Clin}, 74:229--263, 2024.

\bibitem{zhai2024}
Wei Zhai.
\newblock Technical advances in gastrointestinal endoscopy in the diagnosis of early gastric cancer.
\newblock {\em Int J Public Health Med Res}, 1:11--17, 2024.

\bibitem{ansari2023}
Yahya Ansari, Omar Mourad, Khalid Qaraqe, and Erchin Serpedin.
\newblock Deep learning for ecg arrhythmia detection and classification: an overview of progress for period 2017-2023.
\newblock {\em Front Physiol}, 14:1246746, 2023.

\bibitem{ulhaq2024}
Ehtisham Ul~Haq, Qi~Yong, Zhang Yuan, Huang Jianjun, Rizwan Ul~Haq, and Xing Qin.
\newblock Accurate multiclassification and segmentation of gastric cancer based on a hybrid cascaded deep learning model with a vision transformer from endoscopic images.
\newblock {\em Information Sciences}, 670:120568, 2024.

\bibitem{mizumoto2018}
Toshihiko Mizumoto, Toru Hiyama, Duc~T. Quach, Yoji Sanomura, Yoji Urabe, Shiro Oka, Koji Arihiro, Shinji Tanaka, and Kazuaki Chayama.
\newblock Magnifying endoscopy with narrow band imaging in estimating the invasion depth of superficial esophageal squamous cell carcinomas.
\newblock {\em Digestion}, 98(4):249--256, 2018.

\bibitem{muto2016}
Manabu Muto, Kenshi Yao, Mitsuru Kaise, Minoru Kato, Noriya Uedo, Koichi Yagi, and Hisao Tajiri.
\newblock Magnifying endoscopy simple diagnostic algorithm for early gastric cancer (mesda-g).
\newblock {\em Digestive Endoscopy}, 28(4):379--393, 2016.

\bibitem{wang2019}
Lei Wang, Zheng Liu, Huibo Kou, Hongsheng He, Bin Zheng, Lin Zhou, and Ying Yang.
\newblock Double contrast-enhanced ultrasonography in preoperative t staging of gastric cancer: A comparison with endoscopic ultrasonography.
\newblock {\em Frontiers in Oncology}, 9(66), 2019.

\bibitem{osawa2012}
Hiroyuki Osawa, Hironori Yamamoto, Yoshinobu Miura, Makoto Yoshizawa, Keijiro Sunada, Kiichi Satoh, and Kentaro Sugano.
\newblock Diagnosis of extent of early gastric cancer using flexible spectral imaging color enhancement.
\newblock {\em World Journal of Gastrointestinal Endoscopy}, 4(8):356--361, 2012.

\bibitem{teng2024}
Fei Teng, Yanfeng Fu, Ailing Wu, Yutian Xian, Jian Lin, Rui Han, and Yifei Yin.
\newblock Computed tomography-based predictive model for the probability of lymph node metastasis in gastric cancer: A meta-analysis.
\newblock {\em Journal of Computer Assisted Tomography}, 48(1):19--25, 2024.

\bibitem{ma2023}
Lin Ma, Xin Su, Liang Ma, Xia Gao, and Meng Sun.
\newblock Deep learning for classification and localization of early gastric cancer in endoscopic images.
\newblock {\em Biomedical Signal Processing and Control}, 79:104200, 2023.

\bibitem{urakawa2023}
Shogo Urakawa, Takashi Michiura, Shingo Tokuyama, Yoshihiro Fukuda, Yasuhiro Miyazaki, Norito Hayashi, and Kazuki Yamabe.
\newblock Preoperative diagnosis of tumor depth in gastric cancer using transabdominal ultrasonography compared to using endoscopy and computed tomography.
\newblock {\em Surgical Endoscopy}, 37(5):3807--3813, 2023.

\bibitem{wu2023}
Aibo Wu, Chengru Wu, Qing Zeng, Yizheng Cao, Xiaogang Shu, Lin Luo, Zhengyu Feng, Yuanping Tu, Zhuang Jie, Yiping Zhu, et~al.
\newblock Development and validation of a ct radiomics and clinical feature model to predict omental metastases for locally advanced gastric cancer.
\newblock {\em Scientific Reports}, 13(1):8442, 2023.

\bibitem{saito2024}
Yutaka Saito.
\newblock Top tips for performing high-quality optical zoom chromocolonoscopy (with video).
\newblock {\em Gastrointestinal Endoscopy}, 100(1):122--127, 2024.

\bibitem{umegaki2023}
Eiji Umegaki, Hiroaki Misawa, Osamu Handa, Hiroshi Matsumoto, and Akiko Shiotani.
\newblock Linked color imaging for stomach.
\newblock {\em Diagnostics (Basel)}, 13(3):467, 2023.

\bibitem{cho2024}
Hyunwoo Cho, Daeyoung Moon, Seung~Min Heo, Junhee Chu, Hyeonju Bae, Sunghoon Choi, Yeonseung Lee, Donghyun Kim, Yungho Jo, Kyungmin Kim, and et~al.
\newblock Artificial intelligence-based real-time histopathology of gastric cancer using confocal laser endomicroscopy.
\newblock {\em npj Precision Oncology}, 8(1):131, 2024.

\bibitem{chen2024}
Zhenyu Chen, Shuang Fu, Ming Li, Wei Zhang, and Haibo Ou.
\newblock Exploring artificial neural network combined with laser-induced auto-fluorescence technology for noninvasive in vivo upper gastrointestinal tract cancer early diagnosis.
\newblock {\em IJS Oncology}, 5(0):e831--e837, 2024.

\bibitem{srivastava2019}
Sudhir Srivastava, Eugene~J. Koay, Alexander~D. Borowsky, Angelo~M. De~Marzo, Sharmistha Ghosh, Peter~D. Wagner, and Barnett~S. Kramer.
\newblock Cancer overdiagnosis: a biological challenge and clinical dilemma.
\newblock {\em Nature Reviews Cancer}, 19(6):349--358, 2019.

\bibitem{jamil2021}
Danish Jamil.
\newblock Diagnosis of gastric cancer using machine learning techniques in healthcare sector.
\newblock {\em Informatica (Ljubljana)}, 45(7):147--166, 2021.
\newblock URN:NBN:SI:DOC-YA1Q4FKB.

\bibitem{yalamarthi2004}
S.~Yalamarthi, P.~Witherspoon, D.~McCole, and C.~D. Auld.
\newblock Missed diagnoses in patients with upper gastrointestinal cancers.
\newblock {\em Endoscopy}, 36(10):874--879, 2004.

\bibitem{macdonald2006}
S.~Macdonald, U.~Macleod, N.~C. Campbell, D.~Weller, and E.~Mitchell.
\newblock Systematic review of factors influencing patient and practitioner delay in diagnosis of upper gastrointestinal cancer.
\newblock {\em British Journal of Cancer}, 94(9):1272--1280, 2006.

\bibitem{ikenoyama2021}
Y.~Ikenoyama, T.~Hirasawa, M.~Ishioka, K.~Namikawa, S.~Yoshimizu, Y.~Horiuchi, A.~Ishiyama, T.~Yoshio, T.~Tsuchida, Y.~Takeuchi, S.~Shichijo, N.~Katayama, J.~Fujisaki, and T.~Tada.
\newblock Detecting early gastric cancer: Comparison between the diagnostic ability of convolutional neural networks and endoscopists.
\newblock {\em Digestive Endoscopy}, 33(1):141--150, 2021.

\bibitem{yoon2019}
H.~J. Yoon, S.~Kim, J.~H. Kim, J.~S. Keum, S.~I. Oh, J.~Jo, J.~Chun, Y.~H. Youn, H.~Park, I.~G. Kwon, S.~H. Choi, and S.~H. Noh.
\newblock A lesion-based convolutional neural network improves endoscopic detection and depth prediction of early gastric cancer.
\newblock {\em Journal of Clinical Medicine}, 8(9):1310, 2019.

\bibitem{tang2020}
D.~Tang, L.~Wang, T.~Ling, Y.~Lv, M.~Ni, Q.~Zhan, Y.~Fu, D.~Zhuang, H.~Guo, X.~Dou, W.~Zhang, G.~Xu, and X.~Zou.
\newblock Development and validation of a real-time artificial intelligence-assisted system for detecting early gastric cancer: A multicentre retrospective diagnostic study.
\newblock {\em EBioMedicine}, 62:103146, 2020.

\bibitem{wu2019}
L.~Wu, W.~Zhou, X.~Wan, J.~Zhang, L.~Shen, S.~Hu, Q.~Ding, G.~Mu, A.~Yin, X.~Huang, J.~Liu, X.~Jiang, Z.~Wang, Y.~Deng, M.~Liu, R.~Lin, T.~Ling, P.~Li, Q.~Wu, P.~Jin, J.~Chen, and H.~Yu.
\newblock A deep neural network improves endoscopic detection of early gastric cancer without blind spots.
\newblock {\em Endoscopy}, 51(6):522--531, 2019.

\bibitem{shibata2020}
T.~Shibata, A.~Teramoto, H.~Yamada, N.~Ohmiya, K.~Saito, and H.~Fujita.
\newblock Automated detection and segmentation of early gastric cancer from endoscopic images using mask r-cnn.
\newblock {\em Appl. Sci.}, 10:3842, 2020.

\bibitem{goyal2020}
S.~Goyal, A.~Raghunathan, M.~Jain, H.~Simhadri, and P.~Jain.
\newblock Drocc: deep robust one-class classification.
\newblock In {\em Proceedings of the 37th International Conference on Machine Learning}, 2020.

\bibitem{Shahid2015}
N.~Shahid, I.~H. Naqvi, and S.~B. Qaisar.
\newblock One-class support vector machines: analysis of outlier detection for wireless sensor networks in harsh environments.
\newblock {\em Artificial Intelligence Review}, 43(4):515--563, 2015.

\bibitem{wenzhu2019}
S.~Wenzhu, H.~Wenting, X.~Zufeng, and C.~Jianping.
\newblock Overview of one-class classification.
\newblock In {\em 2019 IEEE 4th International Conference on Signal and Image Processing (ICSIP)}, pages 6--10, 2019.

\bibitem{yessoufou2023}
F.~Yessoufou and J.~Zhu.
\newblock One-class convolutional neural network (oc-cnn) model for rapid bridge damage detection using bridge response data.
\newblock {\em KSCE Journal of Civil Engineering}, 27(4):1640--1660, 2023.

\bibitem{laGrassa2022}
R.~La~Grassa, I.~Gallo, and N.~Landro.
\newblock Ocmst: One-class novelty detection using convolutional neural network and minimum spanning trees.
\newblock {\em Pattern Recognition Letters}, 155:114--120, 2022.

\bibitem{costanzoS2021}
A.~Costanzo, V.~Loscri, and M.~Biagi.
\newblock Adaptive modulation control for visible light communication systems.
\newblock {\em Journal of Lightwave Technology}, 39(9):2780--2789, 2021.

\bibitem{zhu2016}
L.~Zhu, J.~Qin, J.~Wang, T.~Guo, Z.~Wang, and J.~Yang.
\newblock Early gastric cancer: Current advances of endoscopic diagnosis and treatment.
\newblock {\em Gastroenterology Research and Practice}, 2016:9638041, 2016.
\newblock Epub 2016 Jan 17. PMID: 26884753; PMCID: PMC4739216.

\bibitem{mejia2017}
L.~K. Mejía-Pérez, S.~Abe, T.~Stevens, M.~A. Parsi, S.~N. Jang, I.~Oda, J.~J. Vargo, Y.~Saito, and A.~Bhatt.
\newblock A minimally invasive treatment for early gi cancers.
\newblock {\em Cleveland Clinic Journal of Medicine}, 84(9):707--717, 2017.
\newblock PMID: 28885903.

\bibitem{MaS2023}
S.~Ma, M.~Zhou, Y.~Xu, X.~Gu, M.~Zou, G.~Abudushalamu, Y.~Yao, X.~Fan, and G.~Wu.
\newblock Clinical application and detection techniques of liquid biopsy in gastric cancer.
\newblock {\em Molecular Cancer}, 22(1):7, 2023.
\newblock PMID: 36627698; PMCID: PMC9832643.

\bibitem{sampath2021}
V.~Sampath, I.~Maurtua, J.~J. Aguilar~Martín, and A.~Gutierrez.
\newblock A survey on generative adversarial networks for imbalance problems in computer vision tasks.
\newblock {\em Journal of Big Data}, 8(1):27, 2021.
\newblock Epub 2021 Jan 29; PMID: 33552840; PMCID: PMC7845583.

\end{thebibliography}
\bibliographystyle{unsrt}

\section*{Supporting Information}
The semi-supervised GS-optimized search Goal Generation reduce the number of mis-classification and recognition percentage of true lesions to the pseudo lesion are necessary and this is further subject. \\

The optimization of hyperparameters in this method is a crucial step in developing an efficient object detection model through DL methods, which are easy to use. Optimization can improve the overall performance, prediction accuracy, and generalization capacity of models, particularly when they are used to predict unseen data. We can see that the batch size of 5 and 50 epochs achieved the best result of about 10\% accuracy.The optimized parameters, their search spaces, and their determined optimal values are shown in \textbf{Supplementary Table~\ref{app1} \& Supplementary Table~ \ref{app2}}.\\

The search for the optimum 2 variables has the search range of [0 1] and [0 1] with a divisions vector of [min(gm) max(gm)] and a minimum range vector of [mean((unique(gm)))
Max(gm)] or  [min(gm) mean((unique(gm)))] . The search employs a maximum of 50 iterations and a function tolerance of max(variables value):
\setcounter{figure}{0}
\renewcommand{\thefigure}{S\arabic{figure}}


\begin{sidewaystable}
\caption{Collection of best multirate location on OCT-X learning}\label{app1}
\begin{tabularx}{\textheight}{lXXXXXXXXX}
\toprule
Iteration & \multicolumn{3}{c}{PM1} & \multicolumn{3}{c}{CM1} & \multicolumn{3}{c}{CM2} \\
\cmidrule(lr){2-4} \cmidrule(lr){5-7} \cmidrule(lr){8-10}
 & \makebox[2cm]{Peak Value} & \makebox[2cm]{Low Retrieval} & \makebox[2cm]{High Retrieval} & \makebox[2cm]{Peak Value} & \makebox[2cm]{Low Retrieval} & \makebox[2cm]{High Retrieval} & \makebox[2cm]{Peak Value} & \makebox[2cm]{Low Retrieval} & \makebox[2cm]{High Retrieval} \\
\midrule
Optimized 1 & 1 & 0.965816 & 0.967216 & 1 & 0.962257 & 0.962657 & 1 & 0.959431 & 0.961631 \\
Optimized 2 & 2 & 0.961216 & 0.963416 & 2 & 0.962582 & 0.962782 & 2 & 0.950418 & 0.960218 \\
Optimized 3 & 3 & 0.954898 & 0.964898 & 3 & 0.964013 & 0.966213 &  & 0.950318 & 0.960318 \\
Optimized 4 & 4 & 0.956097 & 0.961297 & 4 & 0.965621 & 0.965821 & 3 & 0.95205 & 0.95905 \\
Optimized 5 & 5 & 0.965164 & 0.966964 & 5 & 0.965865 & 0.966065 &  & 0.95195 & 0.95915 \\
Optimized 6 & 6 & 0.959867 & 0.963667 & 6 & 0.964812 & 0.966412 &  & 0.95185 & 0.95925 \\
Optimized 7 &  & 0.959767 & 0.963767 & 7 & 0.965615 & 0.966015 & 4 & 0.955983 & 0.957983 \\
Optimized 8 & 7 & 0.96559 & 0.96619 & 8 & 0.963678 & 0.964878 & 5 & 0.954674 & 0.958274 \\
Optimized 9 & 8 & 0.963432 & 0.965032 & 9 & 0.963472 & 0.964672 & 6 & 0.954568 & 0.959368 \\
Optimized 10 & 9 & 0.958928 & 0.963328 & 10 & 0.963658 & 0.964858 &  & 0.953668 & 0.960268 \\
Optimized 11 & 10 & 0.958514 & 0.968314 & 11 & 0.963939 & 0.964939 &  & 0.953568 & 0.960368 \\
Optimized 12 &  & 0.958414 & 0.968414 & 12 & 0.963787 & 0.964787 & 7 & 0.952669 & 0.956069 \\
Optimized 13 & 11 & 0.960664 & 0.961264 & 13 & 0.963956 & 0.964556 & 8 & 0.954921 & 0.957921 \\
\bottomrule
\end{tabularx}
{\footnotesize{\textbf{Note: }This table presents the iteration search results for PM1, CM1, and CM2, including peak values of thresholds, low threshold retrieval, and high threshold retrieval. Due to space constraints, only the top two benchmarking models and our proposed model are presented.}}
\end{sidewaystable}

\begin{sidewaystable} %
\centering
\caption{Collection of best multirate location on OCT-X learning (CONT.)}\label{app2}
\begin{tabularx}{\textheight}{lXXXXXXXXX}%
\toprule
Iteration & \multicolumn{3}{c}{PM1} & \multicolumn{3}{c}{CM1} & \multicolumn{3}{c}{CM2} \\
\cmidrule(lr){2-4} \cmidrule(lr){5-7} \cmidrule(lr){8-10}
 & \makebox[2cm]{Peak Value} & \makebox[2cm]{Low Retrieval} & \makebox[2cm]{High Retrieval} & \makebox[2cm]{Peak Value} & \makebox[2cm]{Low Retrieval} & \makebox[2cm]{High Retrieval} & \makebox[2cm]{Peak Value} & \makebox[2cm]{Low Retrieval} & \makebox[2cm]{High Retrieval} \\
\midrule
Optimized 14 & 12 & 0.959026 & 0.967826 & 14 & 0.963515 & 0.964715 & 9 & 0.959644 & 0.959844 \\
Optimized 15 &  & 0.958926 & 0.967926 & 15 & 0.965667 & 0.966267 &  & 0.958144 & 0.961344 \\
Optimized 16 & 13 & 0.960902 & 0.966302 & 16 & 0.965628 & 0.967428 & 10 & 0.955969 & 0.958969 \\
Optimized 17 &  & 0.960802 & 0.966402 &  & 0.965328 & 0.967728 & 11 & 0.957171 & 0.958971 \\
Optimized 18 & 14 & 0.963924 & 0.966924 & 17 & 0.962885 & 0.964085 & 12 & 0.957094 & 0.958694 \\
Optimized 19 & 15 & 0.959281 & 0.963481 & 18 & 0.96335 & 0.9645 & 13 & 0.957689 & 0.962289 \\
Optimized 20 & 16 & 0.959917 & 0.964717 & 19 & 0.962779 & 0.964179 & 14 & 0.956046 & 0.960246 \\
Optimized 21 &  & 0.959817 & 0.964817 & 20 & 0.963178 & 0.964978 & 15 & 0.956613 & 0.960213 \\
Optimized 22 & 17 & 0.964569 & 0.965369 & / & / & / & 16 & 0.956206 & 0.959806 \\
Optimized 23 & 18 & 0.96612 & 0.96732 & / & / & / & 17 & 0.956252 & 0.959852 \\
Optimized 24 & 19 & 0.958251 & 0.962851 & / & / & / & 18 & 0.955839 & 0.960439 \\
Optimized 25 & 20 & 0.962434 & 0.964234 & / & / & / & 19 & 0.955229 & 0.957629 \\
Optimized 26 & / & / & / & / & / & / & 20 & 0.956602 & 0.959402 \\
\hline
\end{tabularx}
{\footnotesize{\textbf{Note: }Results are shown for iterations Optimized 1 through Optimized 26 for PM1, CM1, and CM2 models including peak values of thresholds, low threshold retrieval, and high threshold retrieval. Due to space constraints, only the top two benchmarking models and our proposed model are presented. }}
\end{sidewaystable}

\label{example_video}
 \begin{table}
 \begin{adjustwidth}{-1.5in}{0in}
\centering
\caption{Optimized param search of FDT-GS agent}
\begin{tabularx}{\linewidth}{lXXX}
\toprule
\textbf{Round} & \textbf{Performance} & \textbf{Low Threshold Retrieval} & \textbf{High Threshold Retrieval} \\
\midrule
1 & 1.5 & 0.483 & 0.97 \\
2 & 0.5 & 0.4829 & 0.971 \\
3 & 4.5 & 0.4827 & 0.973 \\
4 & 5 & 0.4827 & 0.973 \\
5 & 5.5 & 0.4831 & 0.969 \\
6 & 5.5 & 0.4826 & 0.974 \\
7 & 6.4 & 0.4824 & 0.976 \\
8 & 6.5 & 0.4823 & 0.977 \\
\midrule
\textbf{AVG} & / & \textbf{0.4827} & \textbf{0.9729} \\
\bottomrule
\end{tabularx}
\label{app3}
\justifying{
{\footnotesize{\textbf{Note:} Iteration search was conducted with initial parameters $<0.5$ and $>0.8$, with an echo of 200. The table shows the best performance of the iteration in every round.}}}
\end{adjustwidth}

\end{table}

\begin{table}
\begin{adjustwidth}{-1.5in}{0in}
\centering
\caption{Optimized param search of FDT-GS agent (CONT.)}

\begin{tabularx}{\linewidth}{lXXX}
\toprule
\textbf{Round} & \textbf{Performance} & \textbf{Low Threshold Retrieval} & \textbf{High Threshold Retrieval} \\
\midrule
1 & 1.5 & 0.483 & 0.97 \\
2 & 0.5 & 0.4524 & 0.976 \\
3 & 4.5 & 0.4525 & 0.975 \\
4 & 5 & 0.4523 & 0.977 \\
5 & 5.5 & 0.4526 & 0.974 \\
\midrule
\textbf{AVG} & / & \textbf{0.4525} & \textbf{0.9755} \\
\bottomrule
\end{tabularx}
\label{app4}
\justifying{
{\footnotesize\textbf{Note:} Iteration search was conducted with initial parameters $<0.5$ and $>0.5$, with an echo of 500. The table shows the best performance of the iteration in every round.}}
\end{adjustwidth}

\end{table}

\end{document}